\documentclass[reprint,twocolumn,aps,prl,nopacs,superscriptaddress,longbibliography]
{revtex4-2}
\usepackage{float}
\usepackage[english]{babel}
\usepackage[utf8]{inputenc}
\usepackage{fancyhdr}
\usepackage{sidecap}
\usepackage{multirow}
\usepackage{tabularx}

\usepackage{graphicx}
\usepackage{braket}
\usepackage{tikz}
\usepackage{color}
\usepackage{amsmath}
\usepackage[colorlinks=True, linkcolor=blue, filecolor=magenta, urlcolor=blue,citecolor=blue]{hyperref}

\usepackage{graphicx}
\usepackage{comment}
\usepackage{multirow}
\usepackage{xspace}

\newcommand{\subm}[1]{_{\mathrm {#1}}}

\usepackage{xcolor}

\def\m1{{^{-1}}}

\begin{document}
\title{\Large{\textbf {Weak-anti-localization-to-spin-dependent scattering at a proximity-magnetized heavy metal interface}}}

\author{Hisakazu Matsuki}
\affiliation{Department of Materials Science $\&$ Metallurgy, University of Cambridge, 27 Charles Babbage Road, Cambridge CB3 0FS, United Kingdom}
\affiliation{Institute for Chemical Research, Kyoto University, Uji, Kyoto 611-0011, Japan}
\affiliation{Toyota Riken–Kyoto University Research Center (TRiKUC), Kyoto University, Kyoto 606-8501, Japan}

\author{Guang Yang}
 \email{gy251@buaa.edu.cn}
\affiliation{School of Integrated Circuit Science and Engineering, Beihang University, Beijing 100191, China}

\author{Jiahui Xu}
\affiliation{Department of Materials Science $\&$ Metallurgy, University of Cambridge, 27 Charles Babbage Road, Cambridge CB3 0FS, United Kingdom}

\author{Vitaly N. Golovach}
\email{vitaly.golovach@ehu.eus}
\affiliation{Centro de Física de Materiales (CFM-MPC) Centro Mixto CSIC-UPV/EHU, E-20018 Donostia-San Sebastián, Spain}
\affiliation{Donostia International Physics Center (DIPC), 20018 Donostia- San Sebastián, Spain}
\affiliation{IKERBASQUE, Basque Foundation for Science, 48013 Bilbao, Basque Country, Spain}

\author{Yu He}
\affiliation{School of Integrated Circuit Science and Engineering, Beihang University, Beijing 100191, China}

\author{Jiaxu Li}
\affiliation{School of Integrated Circuit Science and Engineering, Beihang University, Beijing 100191, China}

\author{Alberto Hijano}
\affiliation{Centro de Física de Materiales (CFM-MPC) Centro Mixto CSIC-UPV/EHU, E-20018 Donostia-San Sebastián, Spain}
\affiliation{Department of Condensed Matter Physics, University of the Basque Country UPV/EHU, 48080 Bilbao, Spain}
\affiliation{Department of Physics and Nanoscience Center, University of Jyväskylä, P.O. Box 35 (YFL), FI-40014 University of Jyväskylä, Finland}

\author{Niladri Banerjee}
\affiliation{Department of Physics, Blackett Laboratory, Imperial College London, London SW7 2AZ, United Kingdom}

\author{Iuliia Alekhina}
\affiliation{Department of Materials Science $\&$ Metallurgy, University of Cambridge, 27 Charles Babbage Road, Cambridge CB3 0FS, United Kingdom}

\author{Nadia Stelmashenko}
\affiliation{Department of Materials Science $\&$ Metallurgy, University of Cambridge, 27 Charles Babbage Road, Cambridge CB3 0FS, United Kingdom}

\author{F. Sebastian Bergeret}
\affiliation{Centro de Física de Materiales (CFM-MPC) Centro Mixto CSIC-UPV/EHU, E-20018 Donostia-San Sebastián, Spain}
\affiliation{Donostia International Physics Center (DIPC), 20018 Donostia- San Sebastián, Spain}

\author{Jason W. A. Robinson}
 \email{jjr33@cam.ac.uk}
\affiliation{Department of Materials Science $\&$ Metallurgy, University of Cambridge, 27 Charles Babbage Road, Cambridge CB3 0FS, United Kingdom}

\begin{abstract}
A change in a material’s electrical resistance with magnetic field (magnetoresistance) results from quantum interference effects and/or spin-dependent transport, depending on materials properties and dimensionality. In disordered conductors, electron interference leads to weak localization or anti-localization; in contrast, ferromagnetic conductors support spin-dependent scattering, leading to giant magnetoresistance (GMR). By varying the thickness of Au between 4 and 28 nm in a EuS/Au/EuS spin-switches, we observe a crossover from weak anti-localization to interfacial GMR. The crossover is related to a magnetic proximity effect in Au due to electron scattering at the insulating EuS interface. The proximity-induced exchange field in Au suppresses weak anti-localization, consistent with Maekawa-Fukuyama theory. With increasing Au thickness, GMR emerges along with spin Hall magnetoresistance. These findings demonstrate spin transport governed by interfacial exchange fields, building a framework for spintronic functionality without metallic magnetism.

\end{abstract}
\maketitle

Electron spin transport at interfaces between magnetic insulators and nonmagnetic metals (MI/N) is central to the development of spintronic and magnonic device technologies. These interfaces have been investigated using spin pumping~\cite{mosendz_quantifying_2010,heinrich_spin_2011,vlietstra_detection_2016}, lateral spin valves~\cite{dejene_control_2015,muduli_detection_2018,das_temperature_2019}, the spin Seebeck effect~\cite{uchida_spin_2010,huang_transport_2012,qu_intrinsic_2013,costa_recent_2022,vlietstra_simultaneous_2014,kumawat_enhanced_2024}, the spin Peltier effect~\cite{flipse_observation_2014,daimon_thermal_2016,ohnuma_theory_2017,kikkawa_cryogenic_2023}, and spin Hall magnetoresistance (SMR)~\cite{chen_theory_2013,zhang_theory_2019,nakayama_spin_2013,althammer_quantitative_2013,meyer_anomalous_2015,mallick_magnetoresistance_2020,gomez-perez_strong_2020,qin_spin_2021,rosenberger_quantifying_2021}. However, only a few experiments report giant magnetoresistance (GMR) in these structures since this effect is associated with interfaces between metallic ferromagnets and nonmagnetic metals ~\cite{wu_magnon_2018}. Examples include GMR in MI/N systems involving III–V magnetic semiconductors~\cite{xiang_noncollinear_2007,chung_magneto-transport_2008,takase_current--plane_2020}.

Quasi-two-dimensional disordered metallic thin films exhibit quantum interference effects at low temperatures, resulting in an increase in electrical resistance. This defines the weakly localized regime (WLR), characterized by coherent electron backscattering in diffusive conductors~\cite{maekawa_magnetoresistance_1981,bergmann_weak_1984}. In the WLR, magnetoresistance (MR), defined as $(R_H-R_{H=0})/R_{H=0}$, is governed by the strength of spin-orbit coupling (SOC) and the orientation of the applied magnetic field ($H$).

For metallic thin films with negligible SOC, an out-of-plane (OOP) magnetic field suppresses quantum interference, leading to negative MR, while an in-plane (IP) field has little effect on MR~\cite{maekawa_magnetoresistance_1981}. In contrast, for strong SOC, MR is observed for both out-of-plane and in-plane fields~\cite{maekawa_magnetoresistance_1981}. The MR response to an out-of-plane field is described by the Hikami-Larkin-Nagaoka (HLN) theory~\cite{hikami_spin-orbit_1980}, while the response to an in-plane field is captured by the Maekawa-Fukuyama (MF) theory~\cite{maekawa_magnetoresistance_1981}, which accounts for the interplay between Zeeman energy ($\frac{1}{2}g\mu\subm{B}\mu_0H$) and SOC.
For isotropic SOC and elastic scattering times shorter than the spin-orbit scattering time ($\tau_0 \ll \tau_{\mathrm{so}}$), MF theory predicts positive MR with an in-plane magnetic field.

While MR measurements in thin-films under out-of-plane magnetic fields follow HLN theory~\cite{matetskiy_weak_2019,ozeri_modification_2023}, experimental studies showing in-plane MR are scarce~\cite{komori_experimental_1981,kawaguti_positive_1982}. To our knowledge, MR from a Zeeman field in the WLR, induced by a magnetic exchange field (MEF) effect, has not been demonstrated in the literature.
At a MI/N interface, the imaginary part of the spin-mixing conductance ($G_i$) gives rise to a proximity MEF, acting as a Zeeman field within the metal layer~\cite{zhang_theory_2019,geng_superconductor-ferromagnet_2023}, given by
\begin{equation}
G_i \approx g\pi G_0 N\subm{F} \mu\subm{B} \mu_0 H\subm{ex} d,
\label{eq1}
\end{equation}
where $g$ is the Land\'e g-factor, $G_0$ is the quantum conductance, $N\subm{F}$ the density of states per spin at the Fermi level, $\mu\subm{B}$ the Bohr magneton, $\mu_0 H\subm{ex}$ the exchange field, and $d$ the film thickness. The interfacial exchange is by $\kappa\subm{int}$, defined via $\kappa\subm{int}=\frac{1}{2}g\mu\subm{B} \mu_0 H\subm{ex}d$. In SMR—which probes spin-dependent conductivity in a heavy metal (HM) modulated by the magnetization of an adjacent ferromagnetic insulator (FI)—$G_i$ generates an anomalous Hall-like signal due to a field-like torque acting on conduction electron spins~\cite{gomez-perez_strong_2020,kosub_anomalous_2018}. SMR thus sensitively probes interfacial exchange fields at FI/HM interfaces.

EuS is a FI with a bulk Curie temperature of 17 K. The large $G_i$ in EuS~\cite{brataas_non-collinear_2006,gomez-perez_strong_2020} leads to a strong interfacial MEF~\cite{wei_strong_2016} in a N metal or a superconductor (S) at an EuS/N or an EuS/S interface, as seen in EuS/Al~\cite{chang_experimental_2013,strambini_revealing_2017,vaitiekenas_zero-bias_2021}, EuS/graphene~\cite{wei_strong_2016}, EuS/Pt~\cite{gomez-perez_strong_2020}, and recently in EuS/Nb and EuS/Au by some of the present authors~\cite{matsuki_realisation_2025}. \\

In this $Letter$, we report Au(3~nm)/EuS(20~nm) /Au($d$)/EuS(10~nm)/SiO$_2$//Si spin-switch structures in which MR is non-monotonic with the thickness $d$ of the central Au layer, indicating a crossover from weak anti-localization in the WLR to GMR. Furthermore, we observe an AH-like signal, originating from SMR. We discuss how the SMR and MR are linked to $G_i$ and hence the MEF effect at the EuS/Au interfaces. Hereafter, we refer to the spin-switches as EuS(20~nm)/Au($d$)/EuS(10~nm).

The spin-switch structures are fabricated on oxidized Si by electron beam evaporation at room temperature, in a chamber with a base vacuum pressure of 1×10$^{-8}$ mbar.
Electrical contact to the Au barrier is made via AlSi pads, ultrasonically bonded through the film stack. A control structure of Au(3~nm)/MgO(2~nm) shows that the 3-nm-thick Au cap has a resistance of about 6000~$\Omega$ at 10 K, about two orders of magnitude larger than the resistance of the 4-nm-thick Au interlayer in the EuS(20~nm)/Au(4~nm)/EuS(10~nm) structure, indicating that in the spin-switch structures, current is confined to the buried layer of Au.

A key result is that the $G_i$ in EuS/Au generates a substantial interfacial MEF, suppressing quantum interference in Au in the thin limit, consistent with MF theory, and enabling GMR. The large $G_i$ is supported by a pronounced AH-like SMR, which is enhanced by the trilayer structure and the long spin-diffusion length in Au.

\begin{figure}[t]
    \begin{center}
    \includegraphics[scale=0.9]{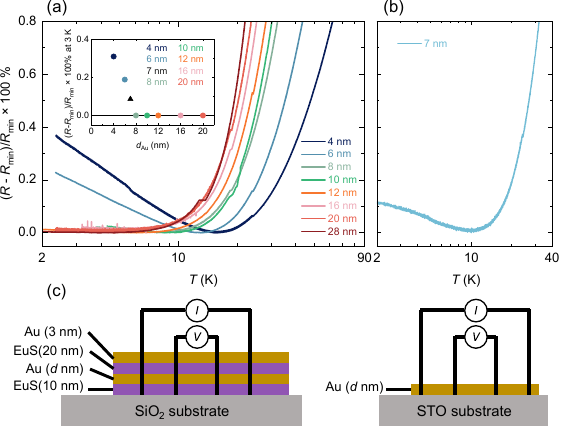}
    \caption{(a) Normalized resistance ($R(T)-R\subm{min})/R\subm{min}$ in zero field cooling vs. temperature ($T$) for EuS(20~nm)/ Au($d$)/EuS(10~nm) and Au(7~nm)/STO control structures. Inset: normalized increase of resistance ($\Delta R$) vs. $d$ at 3 K for spin-switches (circles) and control structures (triangle). (b) $(R(T)-R\subm{min})/R\subm{min}$ in zero-field cooling vs. $T$ for the control structure. (c) Schematic illustrations of the cross-section of the spin-switch and control structure in (a) and (b).}
    \label{fig1}
    \end{center}
\end{figure}

In Fig.~\ref{fig1}(a) we have plotted the normalized resistance $(R(T)-R\subm{min})/R\subm{min}$ versus temperature ($T$) in zero field for the EuS(20~nm)/Au($d$)/EuS(10~nm) structures with different thickness $d$ values (labeled). $R\subm{min}$ is the resistance minima in $R(T)$. The EuS layers are demagnetized. Structures with $d$ = 4 nm and 6 nm show WLR at low temperatures with $R$ minimized between 10 – 20~K, below which $R$ increases ($\Delta R$) by $-$ln($T$). In contrast, spin-switches with $d$ exceeding 8 nm show a monotonic decrease in $R$ with decreasing temperature, inconsistent with WLR. Fig.~\ref{fig1}(b) shows the normalized $R(T)$ of a bare 7-nm-thick Au thin film on single crystal strontium titanate (SrTiO$_3$, STO), grown by magnetron sputtering showing WLR at low temperatures. We note that the identical Au(7~nm)/STO device shows positive MR in an out-of-plane magnetic field as reported in ref.~\cite{ozeri_modification_2023}.
The inset of Fig.~\ref{fig1}(a) shows the normalized increase of resistance of Au at 3 K from $R\subm{min}$. Circles represent the spin-switch structures and the triangle at $d\subm{Au}=7$ nm represents the bare Au thin film in Fig.~\ref{fig1}(b). A linear increase in the normalized $\Delta R$ in structures with the Au thickness of less than 8 nm, with and without EuS, indicating WLR.
The schematic cross-sectional diagram of the structures with AlSi contacts is shown in Fig.~\ref{fig1}(c).

\begin{figure}[t]
    \begin{center}
    \includegraphics[scale=0.8]{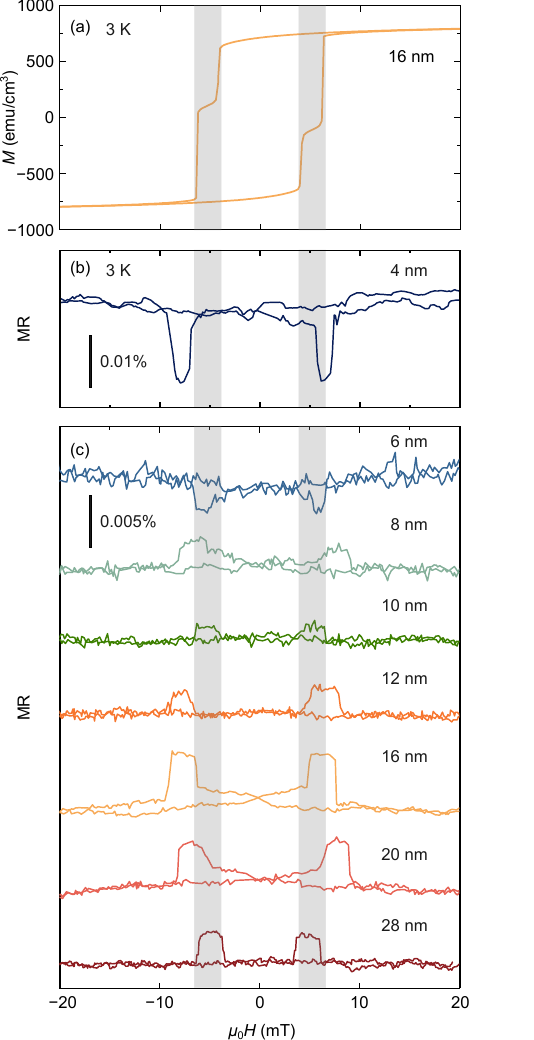}
    \caption{(a) In-plane $M(H)$ for an unpatterned EuS(20~nm)/ Au(16~nm)/EuS(10~nm) structure at 3 K. (b-c) In-plane $R(H)$ of unpatterned EuS(20~nm)/Au($d$)/EuS(10~nm) structures at 3 K. (b): $d=$ 4 nm; (c): $d=$ 6 to 28 nm. Shaded areas indicate AP magnetizations estimated from $M(H)$}
    \label{fig2}
    \end{center}
\end{figure}

We now discuss MR of unpatterned EuS(20~nm)/ Au($d$)/EuS(10~nm) spin-switches, focusing on WLR for $d$ = 4~nm and 6~nm. In Fig.~\ref{fig2}(a) we have plotted the magnetization versus in-plane magnetic field $M(H)$ for $d$ = 16 nm at 3 K, which shows jumps in $M$ at the coercive fields of the two EuS layers ($\pm$7~mT and $\pm$3~mT). By sweeping the magnetic field positive to negative, the relative magnetization-orientation of the two EuS layers switches from parallel (P) to anti-parallel (AP) near $-$3 mT; at $-$7 mT, the magnetization of the magnetically harder EuS layer then switches, recovering the P-state. In Supplementary Fig. 2~\cite{supplement_2025} we have plotted the remanent magnetization for $d$ = 16 nm versus temperature, from which we extrapolate a Curie temperature for the EuS of about 20~K, similar to the $M(T)$ and $M(H)$ behavior reported elsewhere~\cite{gomez-perez_strong_2020,matsuki_realisation_2025}.

\begin{figure}[t]
    \begin{center}
    \includegraphics[scale=1]{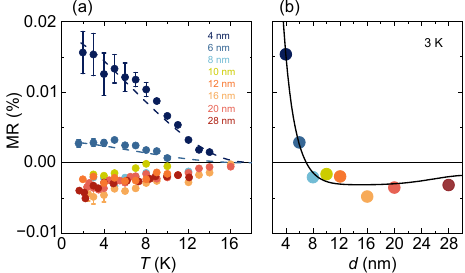}
    \caption{In-plane MR vs. $T$ (a) and vs. $d$ (b) for EuS(20~nm)/Au($d$)/EuS(10~nm). The dotted lines in (a) are fits to Maekawa-Fukuyama theory in the WLR and solid line fits in (b) include a phenomenological effect from WLR and interfacial GMR (see Supplemental Material Part IV in ref~\cite{supplement_2025}).}
    \label{fig3}
    \end{center}
\end{figure}

The effective interfacial MEF induced in N between two FI layers depends on the relative orientation of the FI magnetizations: in the AP-state, the net interfacial MEF in N is minimized whilst in the P-state it should be maximized since the exchange fields at each FI/N interface effectively add for a thin enough layer of N~\cite{miao_spin_2014-1,solis_proximity_2019,takase_current--plane_2020}. 
In this study, we define the magnetoresistance as MR = $(R_{H = 0} - R\subm{AP})/R\subm{AP}\times100\%$, assuming the net interfacial MEF in the AP-state cancels out, where $R_H = 0$ is the resistance at $H = 0$ as the magnetic field is swept from a P-state in the positive field direction to a P-state in the negative field direction. At $H = 0$, the spin-switch is in the P-state, which we define as being equivalent to an isolated N metal in a magnetic field.

In Figs.~\ref{fig2}(b,c), we show in-plane $R(H)$ at 3 K for $d$ = 4 to 28 nm. For $d$ = 4 and 6 nm, resistance minima are seen in the AP-state, consistent with positive MR expected from MF theory. This behavior persists up to the extrapolated Curie temperature of the EuS, as shown in Fig.~\ref{fig3}(a), where the MEF in the P-state vanishes. The temperature-dependent MR of these spin-switches is modeled using MF theory in the WLR, assuming temperature-dependent $\tau_{in}$ and $\kappa\subm{EuS/Au}$, the interfacial MEF at a EuS/Au interface. We use $\kappa\subm{EuS/Au} = 1.5$ meV$\cdot$nm at 2 K from our EuS/Au/Nb/EuS study~\cite{matsuki_realisation_2025}. This parameter follows the $M_r(T)$ curve, with calculations in the Supplemental Material Part~II~\cite{supplement_2025}.

For thicker Au ($d$ = 8 to 28~nm) where WLR diminishes, negative MR is obtained at 3~K [Fig.~\ref{fig2}(c)], persisting up to the Curie temperature of EuS [Fig.~\ref{fig3}(a)]. MR versus $d$ at 3 K [Fig.~\ref{fig3}(b)] is non-monotonic with a sign reversal on a weak suppression of MR up to $d$ = 28 nm. Negative MR indicates GMR, as seen in metallic F/N/F structures where $R\subm{AP} > R\subm{P}$ due to spin-dependent scattering~\cite{baibich_giant_1988,binasch_enhanced_1989} and spin accumulation in N~\cite{valet_theory_1993,fert_nobel_2008}. Although GMR is usually associated with metallic Fs~\cite{wu_magnon_2018}, dilute ferromagnetic semiconductor spin switches have shown GMR near –0.3-0.3$\%$~\cite{xiang_noncollinear_2007,chung_magneto-transport_2008,takase_current--plane_2020}. To our knowledge, GMR in FI/N/FI structures has not been reported. We hypothesize that an interfacial exchange field at the EuS/Au interfaces induces spin-dependent scattering, similar to predictions for EuS/Graphene/EuS~\cite{solis_proximity_2019}. The fit (black line in Fig.~\ref{fig3}(b)) accounts for both weak anti-localization and GMR, with details in the Supplemental Material Part III-V~\cite{supplement_2025}.

\begin{figure}[t]
    \begin{center}
    \includegraphics[scale=0.9]{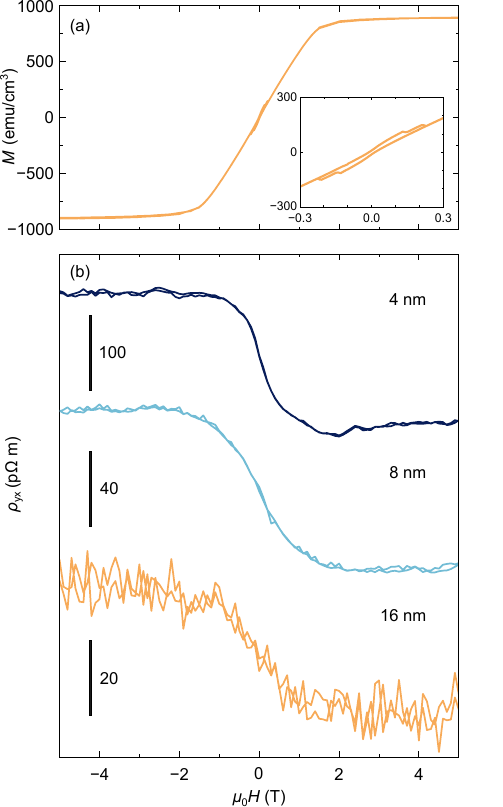}
    \caption{
    (a) Out-of-plane $M(H)$ for an unpatterned EuS(20 nm)/Au($d$)/EuS(10 nm) structure with $d$ = 16 nm at 2 K. Inset: $M(H)$ over a smaller field range.  
    (b) 
    Anomalous Hall resistivity of EuS(20~nm)/Au($d$)/EuS(10~nm) with $d$ = 4, 8 and 16 nm Hall bar structures at 2 K.
    }
    \label{fig4}
    \end{center}
\end{figure}

To investigate further the interfacial exchange field in EuS/Au/EuS, we measured the anomalous Hall effect with an out-of-plane magnetic field. Figure~\ref{fig4} summarizes the main results for different thicknesses of Au.  
The out-of-plane $M(H)$ of an unpatterned $d = 16$ nm structure at 2 K and +5 T to -5 T and back, is shown in Fig.~\ref{fig4}(a). The magnetic moment in an out-of-plane field saturates at about 1.5~T with negligible remanence. The coercive fields of the two EuS layers are about 0.14 T and 0.22 T, as indicated in the inset of Fig.~\ref{fig4}(a).  

A sizable anomalous Hall resistivity $\rho_{yx}$ is obtained for $d = 4$, 8, and 16 nm Hall bars ($100 \times 400~\mu$m$^2$), as shown in Fig.~\ref{fig4}(b) with $H$ scanned from +5 T to -5 T and back, with the trace following the out-of-plane $M(H)$.

We observe $\rho_{yx}$ values of about 90, 42, and 20 p$\Omega\cdot$m for $d = 4$, 8, and 16~nm at 2 K, respectively. The $\rho_{yx}$ value obtained here is three orders of magnitude larger than in a Au(10 nm)/YIG bilayer (0.01 p$\Omega\cdot$m). 
The anomalous Hall resistivity can be attributed to the interplay between SOC in Au and spin-dependent transport generated by the interfacial exchange field. However, from SMR theory~\cite{chen_theory_2013}, the anomalous Hall-like signal for out-of-plane field is expected to vanish. This is because the FI/N/FI trilayers in that work involve materials in which the real part of the spin-mixing conductance ($G_r$) dominates over the imaginary part ($G_i$)~\cite{chen_theory_2013}, which corresponds to most experiments with Y$_3$Fe$_5$O$_{12}$ (YIG)/Pt and related systems~\cite{vlietstra_simultaneous_2014,nakayama_spin_2013,althammer_quantitative_2013,meyer_anomalous_2015,putter_impact_2017,vlietstra_spin-hall_2013,vlietstra_exchange_2013,miao_physical_2014,lu_hybrid_2013,marmion_temperature_2014,wu_strain-modulated_2022,kosub_anomalous_2018}. As shown in Ref.~\cite{chen_theory_2013}, the anomalous Hall-like signal in that case is weak, consistent with YIG/Pt~\cite{kosub_anomalous_2018}.  
In contrast, the strong interfacial exchange field at the Au/EuS interface~\cite{chen_theory_2013} leads to $G_i$ exceeding $G_r$~\cite{gomez-perez_strong_2020,zhang_theory_2019}. In this case, an extension of the original SMR theory~\cite{chen_theory_2013} for a FI/HM/FI structure provides a finite $\rho_{yx}$ in terms of $G_i$:
\begin{equation}
\Delta\rho_{yx}
=-\frac{\theta\subm{SH}^2}{d}
\frac{G_i}{G_i^2+\left[G_r+
\frac{\sigma_0}{2\lambda}\coth\left(\frac{d}{2\lambda}\right)\right]^2},
\label{eq2}\\
\end{equation}
\noindent where $\theta\subm{SH}$ is the spin-Hall angle characterizing the SOC strength, $G_r$ is the real part of the spin-mixing conductance ($G_{\uparrow\downarrow}=G_r+i G_i$),  $\lambda$ is the spin diffusion length, and $\sigma_0$ is the Drude conductivity of the HM layer. We assume both interfaces are similar ($G_{\uparrow\downarrow}^L=G_{\uparrow\downarrow}^R\equiv G_{\uparrow\downarrow}$) and the magnetic configuration is parallel. In the antiparallel state, $\rho_{yx}$ vanishes.
In the present case, the conductivity is large, 
and hence, $\Delta \rho_{yx}$ is proportional to $\theta\subm{SH}^2 G_i$, which, from Eq.~(1), is proportional to the interfacial exchange field. In addition to Eq.~(\ref{eq2}), there is a further contribution to $\Delta \rho_{yx}$ that is proportional to this exchange field. The proper anomalous Hall resistivity~\cite{zhang_nonlocal_2016} scales as $\sim \theta\subm{SH} P$, where $P = (\sigma_\uparrow-\sigma_\downarrow)/(\sigma_\uparrow+\sigma_\downarrow)$ denotes the effective polarization of the conduction electrons in Au induced by the interfacial exchange field, which is also responsible for the GMR seen in thicker Au layers in Fig.~\ref{fig2}.  
Both the generalized SMR expression, Eq.~(2), and this additional contribution account for the sizable anomalous Hall signal in Fig.~\ref{fig4}(b), confirming the key role of interfacial exchange in EuS/Au structures.


In summary, we have observed a sign reversal in the Au-thickness dependence of in-plane MR in EuS/Au/EuS spin-switches, indicating a transition from WLR to GMR. We argue that this is driven by a strong interfacial magnetic exchange field at the EuS/Au interfaces. A pronounced anomalous Hall-like SMR, linked to a large imaginary spin-mixing conductance ($G_i$), supports this interpretation. These results highlight the potential of EuS/Au/EuS trilayers as spintronic elements for efficient spin transport and manipulation.\\

\begin{acknowledgments}
$Acknowledgements-$J.W.A.R. and H.M. acknowledge funding from the EPSRC (EP/P024947/1, EP/R00661X/1, EP/P026311/1, EP/N017242/1).
H.M. also acknowledges support from the Kyoto University Foundation.
G.Y. acknowledges funding from the National Key R\&D Program of China (2022YFA1402604), the Natural Science Foundation of China (52201200), and the Shandong Provincial Natural Science Foundation (ZR2024MA054).
A.H. acknowledges funding from the University of the Basque Country (PIF20/05), the Research Council of Finland via the Finnish Quantum Flagship (359240), and the European Research Executive Agency (101202316). F.S.B. acknowledges funding Spanish MCIN/AEI/10.13039/501100011033 through PID2023-148225NB-C31 and TED2021-130292B-C42, the Basque Government (IT-1591-22), and Horizon Europe (101130224, JOSEPHINE).

\end{acknowledgments}

\bibliography{hisa_zotero_PRL}
\bibliographystyle{apsrev4-2}
\end{document}